\documentclass[preprint]{elsarticle}

\usepackage{amsmath, amsthm, amssymb}
\usepackage{multirow, graphics, graphicx, lscape}

\begin{document}
\title{A Survey of Protocols for Intermittently Connected Delay-Tolerant Wireless Sensor Networks}

\author[unh]{Ying Li \corref{cor1}}
\ead{yws2@cs.unh.edu}

\author[unh]{Radim Bartos}
\ead{rbartos@cs.unh.edu}

\cortext[cor1]{Corresponding author. Tel: +1 603 862 3792; fax: +1 603 862 3493}

\address[unh]{Department of Computer Science, \\ University of New Hampshire, Durham, NH 03824}

\begin{abstract} 
Intermittently Connected Delay-Tolerant Wireless Sensor Networks (ICDT-WSNs), a branch of Wireless Sensor Networks (WSNs), have features of WSNs and the intermittent connectivity of Delay-Tolerant Networks (DTNs). The applications of ICDT-WSNs are increasing in recent years, however, the communication protocols suitable for this category of networks often fall short. Most of the existing communication protocols are designed for either WSNs or DTNs and tend to be inadequate for direct use in ICDT-WSNs. This survey summarizes characteristics of ICDT-WSNs and their communication protocol requirements, and examines the communication protocols designed for WSNs and DTNs in recent years from the perspective of ICDT-WSNs. Opportunities for future research in ICDT-WSNs are also outlined.
\end{abstract}


\maketitle
\pagestyle{plain}\thispagestyle{plain} 

\section{Introduction}
\label{Introduction}


Intermittently Connected Delay-Tolerant Wireless Sensor Networks (ICDT-WSNs) are a new branch of Wireless Sensor Networks (WSNs), which have characteristics of WSNs and Delay-Tolerant Networks (DTNs). These characteristics include the limited energy, low computation capability, small storage, narrow bandwidth, short communication range~\cite{one} and the intermittent connectivity that end-to-end paths do not always exist in networks~\cite{two}. These difficulties make the design of communication protocols for ICDT-WSNs a challenging task, although ICDT-WSNs have been commonly used in areas whose development environments are unsafe or even impossible for human to access. Examples of use include wildlife tracking~\cite{three}, assisting submarine location estimation~\cite{four}, solar-powered autonomous underwater vehicle (SAUV) platform for underwater networks~\cite{five}, coal mine structure surveillance~\cite{six} and sandstorm forecast~\cite{seven}.  

Most of the existing protocols cannot be directly employed in ICDT-WSNs, since they are either designed for WSNs or DTNs that do not take all limitations of ICDT-WSNs into consideration. Without reliable, robust and efficient communication protocols, the performance of ICDT-WSNs is degraded resulting in shortened network life time, decreased propagation speed and increased packet loss rate. As a consequence, the development of ICDT-WSN applications is constrained. 

In this article we list the attributes of ICDT-WSNs and the requirements for communication protocols of this category of networks, outline several communication protocols that have been designed in recent years, and evaluate them from the perspective of ICDT-WSNs for improvement opportunities in communication protocols. Some open problems in ICDT-WSNs and possible directions to address these problems are also discussed in this article.

The rest of this article is organized as follows: Section \ref{Background} gives brief introductions to WSNs and DTNs to provide sufficient background for ICDT-WSNs, and introduces ICDT-WSNs in detail. Transport, network, and link layer communication protocol outlines and evaluations are provided in Section \ref{Communication Protocols}. Section \ref{Open Problems} gives out open problems in ICDT-WSNs and provides possible solutions. Conclusions are drawn in Section \ref{Conclusion}.  

\section{Background}
\label{Background}
\subsection{Wireless Sensor Networks}
\label{Wireless Sensor Networks}

\emph{Wireless Sensor Networks} (WSNs) have been extensively studied and widely used in the recent decade. A WSN can consist of one to several types of sensor nodes such as visual, thermal, acoustic, infrared, radar, low sampling rate magnetic, and seismic~\cite{one}. WSNs are mission-oriented: all sensor nodes of a WSN cooperate together to accomplish the mission of the network, such as collecting environmental data from a designated area and tracking an object. According to the environment the WSNs are developed for, WSNs can be categorized into terrestrial, underwater or underground:

\begin{itemize}
\item \emph{Terrestrial} WSNs are developed above ground, and are usually composed of hundreds to thousands of low-cost sensor nodes~\cite{eight}. The terrestrial WSNs can be used for environment sensing and monitoring, industry monitoring~\cite{nine} and surface exploration. Radio Frequency (RF) communication is widely used in terrestrial WSNs. Energy efficiency is very important for terrestrial WSNs, since the power of sensor nodes is very limited even with solar cells.

\item \emph{Underwater} WSNs consist of a variable number of sensors and vehicles that are sparsely deployed under water for oceanographic data collection, pollution monitoring, offshore exploration, disaster prevention, navigation assistance and tactical surveillance applications~\cite{ten}. Instead of RF communication, acoustic communication is preferred in underwater WSNs, because of the high attenuation of RF in acquatic environments~\cite{eleven}. Compared to the terrestrial WSNs, underwater WSNs suffer more severe challenges: longer propagation delay, less bandwidth, more severely impaired channels and non-rechargeable, limited power.

\item \emph{Underground} WSNs comprise of a number of sensor nodes buried underground or placed in coal mines or caves, used to monitor a variety of underground conditions~\cite{six, twelve}. RF communication can be used in underground WSNs~\cite{six}, but the underground environment causes high attenuation of electromagnetic waves. Akyildiz~\cite{twelve} points out that Magnetic Induction (MI) and seismic waves might be better for communication in underground WSNs. In addition to the challenges of underwater WSNs, signal fade is unavoidable in underground WSNs.
\end{itemize}

With the development of micro-electro-mechanical systems (MEMS) technology, sensor nodes have become smaller, lighter, smarter and cheaper. In addition to the main categories of WSNs mentioned above, WSNs are now being used in airplane surveillance~\cite{thirteen} and body sensor networks~\cite{fourteen, fifteen}.

The network infrastructures, sensor nodes and communication protocols can be different from one WSN to another. Because WSNs are mission oriented, the topology design and device selection for a WSN depends on the application for each WSN. 

Generally, WSNs have little or no infrastructure. According to the manner of node deployment, WSNs can be divided into two groups: \emph{ad hoc} WSNs and \emph{pre-planned} WSNs. Ad hoc WSNs have no infrastructure, the sensor nodes are deployed into a field randomly, possibly scattered from an airplane and left unattended. In order to maintain connectivity and detect failures, the protocols and algorithms for ad hoc WSNs should be able to self-organize. The ad hoc nature makes this category of WSNs suitable for disaster relief and operations in inaccessible areas. Pre-planned WSNs, on the contrary, are more structured networks, and can be grouped into wireless mesh networks. Sensor nodes in pre-planned WSNs are placed at particular positions in a pre-planned manner, such that topologies are well designed beforehand. For several examples of typical pre-planned WSNs see underwater WSNs~\cite{ten, eleven} and underground WSNs~\cite{twelve, six}.

According to the mobility of sensor nodes, WSNs can be categorized into \emph{static} WSNs and \emph{mobile} WSNs. WSNs that only consist of non-moving sensor nodes are static WSNs. WSNs containing self propelled sensor nodes are mobile WSNs. Depending on the design of a network, the movement of sensor nodes in a network can be controllable and predictable. This property not only distinguishes mobile WSNs from MANETs, but also provides an advantage for communication protocol design. 




The communication protocols for WSNs can be classified into connection-oriented and disconnection-oriented. The \emph{connection-oriented} protocols assume that a complete path from a source to a destination in a network always exists. But the \emph{disconnection-oriented} protocols assume that a complete path between a source and a destination in the network does not always exist, and can be highly unstable~\cite{sixteen}. The disconnection-oriented protocols also need to be tolerant to the long propagation delay caused by disconnection. Disconnection-oriented protocols are necessary for intermittently connected WSNs, which is discussed in Subsection \ref{Intermittently Connected Delay-Tolerant WSNs}.

\subsection{Delay-tolerant Networks}
\label{Delay-tolerant Networks}

A \emph{Delay-Tolerant Network} (DTN) is an overlay on top of regional networks\footnote[1]{A regional network is a network in which the communication characteristics are homogeneous.}, and provides interoperability between these networks~\cite{seventeen}. DTNs are challenging networks, where the architectures and communication protocols used in traditional networks may operate poorly. The challenges associated with DTNs are intermittent connectivity, long or variable delay, asymmetric data rates, and high error rates.

The Delay-Tolerant Networking Research Group (DTNRG)~\cite{eighteen} discusses the \emph{bundle layer} as the overlay DTN architecture, which not only provides a transparent communication among different regional networks, but also hides the disconnection and delay from the application layer. 

The bundle layer sits between the application layer and the transport layer in the DTN protocol stack. In the bundle layer, the application data is encapsulated into bundles with bundle headers and passed to the transport layer. The bundle layer implements \emph{store-and-forward} message switching to overcome the network interruption, and provides end-to-end reliability across a DTN through custody transfers. Nodes in DTNs have persistent storages to store bundles and support \emph{custody transfers}. 

Custody transfers achieve end-to-end reliability by employing node-to-node retransmission in the bundle layer to prevent data loss and corruption. If a node requires custody transfers, it starts a time-to-acknowledge retransmission timer after sending a bundle to the next node. If the next node accepts the custody, it returns an acknowledgement to the sender. If no acknowledgement is received before the sender's timer expires, the sender retransmits the bundle. If a node supports custody transfers, it must store a bundle until another node accepts custody or the bundle's time-to-live expires. Otherwise a node only needs to store a bundle until the outbound links are available.

The node name in a DTN consists of two parts, the \emph{region ID} and the \emph{entity ID}.  The bundle layer provides transparent communication among different regional networks through the region ID. The routing within a regional network is based on the entity IDs.

\subsection{Intermittently Connected Delay-Tolerant WSNs}
\label{Intermittently Connected Delay-Tolerant WSNs}

\emph {Intermittently Connected Delay-Tolerant WSNs} (ICDT-WSNs) are WSNs with intermittent connectivity. In such networks, the path between a source and a destination can be highly unstable, and may change or break while being discovered. Underwater WSNs, underground WSNs and mobile WSNs (e.g. ZebraNet~\cite{three}) are well known instances of ICDT-WSNs.

The characteristics of ICDT-WSNs are as follows: 

\begin{description}
\item {\emph{Link characteristics:}} intermittent connectivity, long or variable delay, asymmetric data rates and high error rates.
\item {\emph{Node characteristics:}} limited power, low processing capability, minimal storage, short communication range, and low bandwidth.
\end{description}

The intermittent connectivity is a phenomenon caused by some intrinsic features of WSNs. First, due to the limitation on node energy, low-duty-cycle nodes, which are powered off most of the time, are usually used in WSNs for energy conservation. Second, node failures are common in the harsh environments that WSNs are developed in due to power exhaustion, node damage and corruption. Third, mobile nodes are required in some applications, resulting in nodes moving outside of each other's communication range. Finally, signal attenuation plays a role in the intermittent connectivity as well. As a result of the intermittent connection, the propagation delay of ICDT-WSNs can be long or variable, requiring ICDT-WSNs to be delay-tolerant.

The existing communication protocols for DTNs and WSNs may be unsuitable for ICDT-WSNs, since DTN protocols do not take the limitations on node energy, processing capability and storage into account, and many existing WSN protocols assume that the path between a source and a destination always exists. Consequently, communication protocols for ICDT-WSNs need to be:

\begin{enumerate}
\item \emph{Disconnection-oriented}: the path between a source and a destination is not always stable. Communication protocols should not assume a complete path from a source to a destination always exists. 
\item \emph{Energy efficiency}: energy is very limited in ICDT-WSNs. Communication protocols should take communication duration, communication distance, sensing period, computation complexity and frequency of node status switching (power on and off) into consideration.
\item \emph{Storage friendly}: sensor node storage is much smaller than the storage in a server. Storage management and message priority can improve the performance of ICDT-WSN protocols.
\item \emph{Computation simplicity}: sensor node computation capability is not comparable to that of a server. The low computation complexity of communication protocols is even more important in ICDT-WSNs than in other networks.
\end{enumerate}

\section{Communication Protocols}
\label{Communication Protocols}

In this section, we study several transport, network and link layer protocols designed in recent years. In each subsection, we outline the communication protocols of one layer, evaluate the protocols with respect to the applicability to ICDT-WSNs, and summarize the future improvement opportunities for communication protocols for ICDT-WSNs in that layer.

\subsection{Transport Layer}
\label{Transport Layer}

Due to the characteristics of ICDT-WSNs, the traditional transport protocols that are widely used in Internet (TCP, UDP) cannot be applied in ICDT-WSNs directly~\cite{nineteen}. User Datagram Protocol (UDP) would not be an appropriate option for ICDT-WSN applications that require reliable delivery, such as military surveillance. Transmission Control Protocol (TCP) is inefficient in ICDT-WSNs, since TCP is designed for traditional networks where packet loss is mainly due to traffic congestion~\cite{twenty}. But there are many reasons other than traffic congestion causing packet loss in ICDT-WSNs, such as sensor nodes out of memory, signal attenuation and energy exhaustion. Therefore, the congestion control mechanism in TCP fails to determine traffic problems correctly, leading the protocol to perform poorly. Moreover, the end-to-end reliability of TCP has to be managed by source nodes, which can make the limited energy of sensor nodes drain quickly.  
    
Different applications may require different levels of reliability. As a result, transport protocols that can provide multiple levels of loss recovery are more flexible to meet the various requirements for the diversity of applications in ICDT-WSNs. Because of the multi-hop transmission in ICDT-WSNs, congestion is more likely to happen at sensor nodes that are geographically closer to the sink or have a higher probability of moving into the sink's communication range. Therefore, effective congestion control is required to reduce packet loss, save energy, extend network lifetime and enhance throughput.   
 
There are two approaches for loss recovery and congestion control in ICDT-WSNs: end-to-end and hop-by-hop. End-to-end approaches can be thought of as centralized methods. The sink node is usually the end that manages reliability and executes congestion control. The benefits of end-to-end approaches are: 1)~the sink node usually has plenty of energy; 2) the sink node can have a more complete view of traffic in the whole network than a single sensor node. Hop-by-hop approaches can be thought of as decentralized methods. Every node in the network has the responsibility to provide hop level loss recovery and congestion control. Compared to the end-to-end approaches, hop-by-hop congestion control can react faster when a problem is detected and the retransmission distance of hop-by-hop loss recovery is shorter. But hop-by-hop approaches are less flexible than the end-to-end approaches in providing variable reliability levels. Due to the intermittent connectivity, hop-by-hop approaches are more efficient than end-to-end approaches in ICDT-WSNs. 

Most of the existing transport protocols are designed for either WSNs or DTNs and cannot meet all the requirements of ICDT-WSNs. These transport protocols provide either end-to-end reliability or congestion control. To the best of our knowledge, no literature points out that combining protocols together can achieve better performance than applying a single protocol that provides both services. In the rest of this subsection, we focus our study on transport protocols that provide both end-to-end reliability and congestion control. The outlined transport protocols have been proposed in recent years and have some features than can be used to design transport protocols for ICDT-WSNs.
 
\emph{Hop}: Hop~\cite{twenty-one} provides end-to-end reliability and congestion control in a hop-by-hop manner for wireless mesh network. It uses \emph{block}s as the transmission unit and disables the link layer acknowledgement. A block consists of a number of packets. A node $N_i$ transmits a BSYN message to a node $N_{i+1}$ when finished forwarding a block of packets to $N_{i+1}$ to indicate the end of the block. After $N_{i+1}$ receives the BSYN message, a BACK message is sent by $N_{i+1}$ to $N_{i}$ to indicate the lost packets in that block. When $N_{i}$ receives the BACK message, it retransmits any lost packets indicated by the BACK message to $N_{i+1}$. The procedure continues until the block has been completely received by $N_{i+1}$. Then $N_{i+1}$ forwards the block to its next hop $N_{i+2}$. In this manner, Hop not only guarantees hop level reliability, but also reduces the per-packet based acknowledgements by employing per-block based acknowledgements. 

Hop applies virtual retransmission and in-network storage to decreases the number of end-to-end retransmissions when provides end-to-end reliability. Hop assumes that when transmitting a block along a path $P$, the nodes near $P$ might overhear parts of the block. Nodes in Hop cache the received and overheard packets. If a node $A$ on $P$ fails, another node $B$ in the neighborhood of $A$'s upstream node $C$ is selected as the new relay node. Instead of retransmitting the complete block to $C$, virtual retransmission allows $B$ to only send the missing packets of the block to $C$.

Hop uses backpressure to provide hop-by-hop congestion control. In Hop, each node keeps a threshold H, which is the difference between the number of blocks the node receives and the number of blocks it transmits to the next node. After completely receiving H blocks, a node does not respond to the arriving BSYN messages from upstream nodes until it transmits more blocks to its downstream nodes. The next hop to forward blocks is chosen through backpressure to avoid the heavy traffic directions.
 
\emph{Rate-Controlled Reliable Transport (RCRT)}: RCRT~\cite{twenty-two, twenty-three} is an end-to-end reliable transport protocol for WSNs, which provides centralized congestion control at the sink node and supports concurrent flows. In RCRT, the sink node maintains two lists for each flow. One is the list of missing packets, and the other is the list of out-of-order packets. RTRC takes advantage of a NACK-based end-to-end recovery scheme that entries in the missing packet lists are sent back to the sources through Negative Acknowledgements (NACKs). 

There are three distinct logical components in the congestion control mechanism of RCRT: congestion detection, rate adaption and rate allocation. The intuition of the congestion detection is that if the lost packets are recovered quickly enough, the network is not congested. The way to measure the loss recovery time is the lengths of the out-of-order packet lists. Let $r_i$ be the transmission rate of source $i$, $RRT_i$ be the round trip time of source $i$, and $L_i$ be the length of the out-of-order packet list for source $i$ at an observation time. The number of rounds for recovering the lost packets of source $i$ at the observation time is measured:
\begin{equation}
L_{norm,i} = \frac{L_i}{r_iRRT_i}.
\label{eq: rounds_for_loss_recovery}
\end{equation}
\noindent If $L_{norm,i}$ of any list is higher than the higher threshold, the network is congested. If $L_{norm,i}$ of every list is lower than the lower threshold, the network is declared uncongested.  

The rate adaptation of RCRT uses Additive increase/multiplicative decrease (AIMD) on the aggregate rate of all flows observed at the sink. Whenever RCRT detects the network is congested, it decreases the rate:
\begin{equation}
R(t+1) = R(t)M(t)
\label{eq: aimd_decrease}
\end{equation}
\noindent where $R(t)$ is the sum of currently assigned rates $r_i(t)$ for all flows $i$, $M(t)$ is the multiplicative decrease factor (Equation~\ref{eq: decrease_factor}), and $p_i(t)$ is the current delivery ratio of the congested flow from source $i$. 
\begin{equation}
M(t) = \frac{p_i(t)}{2-p_i(t)}
\label{eq: decrease_factor}
\end{equation}
If the network is uncongested, RCRT increases the rate:
\begin{equation} 
R(t+1) = R(t) + A
\label{eq: aimd_increase}
\end{equation} 
\noindent where $A$ is the additive increase factor, a positive constant. 

After adapting the overall rate, the rate allocation component of RCRT assigns rates $r_i(t)$ to each flow, with the sum of all $r_i(t)$ equal to $R(t)$. The prototype of RCRT provides three rate allocation rules: demand-proportional, demand-limited and fair, which makes RCRT flexible in rate allocation. Demand-proportional is set so that each flow $i$ has desired rate $d_i$, and the assigned rate is calculated as follows:
\begin{equation}
r_i(t) = d_i\rho_i(t)
\label{eq: rate_allocate_1}
\end{equation}
\begin{equation}
\rho_i(t) = \frac{R(t)}{D}
\label{eq: rate_allocate_2}
\end{equation}
\noindent where $\rho_i(t)$ is the allocation ratio, and $D$ is the sum of desired rates of all flows. Demand-limited is defined such that $R(t)$ is divided equally among all flows as long as the assigned rate $r_i$ is less than or equal to the desired rate $d_i$. Fair is defined such that $R(t)$ is divided equally among all flows regardless of desired rate.

Unlike RCRT, \emph{Retiring Replicants}~\cite{twenty-four} utilizes AIMD to provide a hop-by-hop congestion control. Retiring Replicants dynamically controls the amount of replicants in a networks by letting each node maintains the drop and replication counters. According to these numbers, nodes decide to increase or decrease the replicants locally. Retiring Replicants enhances the network performance when the underlying routing protocols are flooding based protocols. However, since Retiring Replicants does not provide reliability, the delivery rate cannot be improved significantly.   

\emph{Sensor Transmission Control Protocol (STCP)}: STCP~\cite{nineteen} supports end-to-end reliability for multiple flows and has the ability to provide multiple levels of reliability. Initially, source nodes send session initiation packets to inform the sink of what flows they will send, such as the number of flows, flow type, transmission rate and reliability level. After source nodes receive ACK packets for their initiation packets, they can start to transmit data. 

If a flow is a continuous flow, the sink will know the expected arrival time of the next packet in the flow through the transmission rate sent in the initiation packet. If the sink does not receive a packet within the expected time frame, it transmits a NACK packet with the sequence number of the missing packet in the flow. When the sink receives the next packet, it transmits an ACK packet. The sink keeps a list of transmitted NACK packets to deal with NACK packet loss by periodically checking the list and resending NACK packets until all missing packets have been received. When a missing packet is received, its entry is removed from the NACK packet list. The source nodes retransmit missing packets after receiving NACK packets. 

Each source node keeps a buffer timer to prevent buffer overflow. The buffer timer depends on the transmission rate of packets and the network conditions. When a source node's buffer  timer expires, the node's buffer is cleaned. 

If a flow is an event-driven flow, the sink sends an ACK after receiving each packet. Source nodes cache packets until corresponding ACKs are received, and then remove confirmed packets from their buffers. Each source node keeps a retransmission timer. When the retransmission timer of a source node expires, packets in the node's buffer are retransmitted. 

STCP treats multi-level reliability for continuous flows and event-driven flows differently. For continuous flows, the sink measures the reliability level by the fraction of the packets received successfully. The sink sends NACKs for retransmission only when the reliability level goes below the required level. But for event-driven flows, source nodes calculate the reliability before transmitting packets. Packets are cached at the source nodes if the reliability goes below the required level.   

Congestion control in STCP is through the binary congestion notification bit in the packet header. Each sensor node maintains two thresholds for its buffer $t_{lower}$ and $t_{higher}$. When the buffer size reaches $t_{lower}$, the sensor node sets the congestion notification bit in the packets it forwards with a certain probability. When the buffer size reaches $t_{higher}$, the sensor node sets the congestion notification bit in every packets it forwards. When receiving a packet with congestion bit set, the sink sets the congestion bit in the ACK packet. After receiving an ACK packet with congestion bit set, the source node either adjusts the transmission rate or routes the successive packets along a different path.

\emph{Flush}: Flush~\cite{twenty-five} provides end-to-end reliability and hop-by-hop congestion control for one flow in a network at a time. Data is divided into packets and sent in a pipelined scheme by Flush. The end-to-end reliability is guaranteed through selective negative acknowledgement (NACK). The sink initializes the transmission by sending a request to a target node. A source node sends packets after receiving a request. The sink keeps track of received packets, and after an estimated round-trip-time the sink sends a NACK packet with the sequence numbers of the first three missing packets. This procedure continues until the sink receives all required packets.

Flush uses a hop-by-hop rate control mechanism to avoid congestion. The algorithm follows two policies: 1) a node transmits only when its successor is free from interference; 2) the sending rate of a node cannot exceed the sending rate of its successor. So the sending interval at node $i$ is the maximum value of either the minimum delay $d_i$ or the sending interval of its successor node $D_{i-1}$. The minimum delay is the sum of $\delta_i$ the time node $i$ takes to transmit a packet,  $\delta_{i-1}$ the time node $i-1$ transmits the packet and $f_{i-1}$ the time the packet moves out of the interference range of node $i-1$.  When a node's queue length exceeds a threshold, the node temporarily doubles its transmission time $\delta$ to increase the delay.

\emph{Burst Forwarding}: Burst forwarding~\cite{twenty-six} alone does not provide end-to-end reliability and congestion control, but it provides a way to employ TCP in WSNs. Burst forwarding is a pipelined forwarding protocol over TCP that can provide high throughput and low energy consumption. Burst forwarding supports single flow transmission, and applies multi-channel operation to achieve high throughput.   

Burst forwarding performs \emph{clear channel assessments} (CCA) before starting transfer a burst. If a sender tests the channel is idle, it starts to transfer a burst. If a receiver tests the channel is not idle, it starts to listen. Packets are forwarded in burst. When a receiver receives a packet, it sends an acknowledgement back to the sender. After received an  acknowledgement, a sender keeps forwarding the following packets. If no acknowledgement received by a sender after forwarded a packet, the sender retransmits the packet.

Burst forwarding provides two-level retransmissions: \emph{link-layer retransmission} and \emph{MAC-layer retransmission}, which can reduce the number of end-to-end retransmissions of TCP. Link-layer retransmissions reduce the number of data transfers between the link and transport layers since the messages can be stored in the radio transceiver. Instead of retransmitting a lost packet, the MAC-layer retransmission transmits the lost packet with the new packets in burst after a back-off time.    

\emph{Asymmetric and reliable transport (ART) mechanism}: ART~\cite{twenty-seven} provides different levels of reliability for two different types of flows. Flows in a network are classified into upstream flows, which are from sensors to the sink, and downstream flows, which are from the sink to sensors. ART assumes these two types of flows have different loads and require different levels of reliability. A downstream flow has lighter load but requires higher reliability that all packets of this flow should be received by the sensors. An upstream flow, on the contrary, has heavier load but requires lower reliability that the sink does not have to receive every packet as long as it acquires the event. ART also assumes that congestion seldom happens in downstream flows but frequently occurs in upstream flows. 

ART proposes a centralized, energy-aware sensor classification algorithm to elect a minimal set of nodes that have higher energy levels and covers the whole target area. The nodes in this set are called \emph{essential nodes} (E-nodes), and all the other nodes in the network are called \emph{non-essential nodes} (N-nodes). This algorithm is executed periodically in order to prolong the network lifetime.

\begin{landscape}
\begin{table*}[htbp]
\centering
\small
\caption{Attribute comparison of transport protocols providing both end-to-end reliability and congestion control}
\begin{tabular}{|p{1.3cm}|p{1.0cm}|p{1.3cm}|p{0.8cm}|p{1.7cm}|p{0.6cm}|p{1.6cm}|p{1.7cm}|p{1.7cm}|p{1.4cm}|p{1.4cm}|}
\hline
\textbf{Protocols}& \multicolumn{4}{|c|}{\textbf{Reliability}} & \multicolumn{4}{|c|}{\textbf{Congestion Control}} &\textbf{Energy Control} & \textbf{Buffer Management} \\ [0.5ex]
\cline{2-9}
& \textbf{Level} & \textbf{Direction} & \textbf{Type} & \textbf{Notification} & \textbf{Type} & \textbf{Detection} & \textbf{Notification} & \textbf{Alleviation} &  & \\ [0.5ex]
\hline
\textbf{Hop} & packet & both & virtual retransmission & BSYN/ BACK & hop-by-hop &  backpressure  & N/A & hold blocks & using~block and~BSYN/ BACK~to reduce overheads &  N/A \\[0.5ex]
\hline
\textbf{RCRT} & packet & upstream & end-to-end & NACK & end-to-end & loss recovery time & explicit & adjust rate by AIMD & N/A & N/A\\[0.5ex]
\hline
\textbf{STCP} & event / packet (variable levels) & upstream  & end-to-end & NACK/ACK & end-to-end & queue length & implicit & slow down the rate~or change to another path & N/A & buffer timer to clean buffer for continuous flows \\[0.5ex]
\hline
\textbf{Flush} & packet & upstream & end-to-end & NACK & hop-by-hop & queue length & N/A & slow down by increasing the delay &  N/A & N/A \\[0.5ex]
\hline
\textbf{ART} & event / query (different levels) & both & end-to-end & NACK/ACK & hop-by-hop & service time & explicit & N-nodes temporarily stop~sending & energy-aware sensor classification & N/A\\[0.5ex]
\hline
\end{tabular}
\label{table:transport_protocols}
\end{table*}
\end{landscape}

ART provides reliability by using asymmetric acknowledgement (ACK) and negative acknowledgement (NACK) between E-nodes and sink. In a downstream transfer, the sink sends a set of queries to an E-node. Each query has a sequence number. If the query is the last one in the set, the sink sets the Poll/Final (P/F) bit. When an E-node receives the queries, it checks sequence numbers to detect query loss. When a gap in sequence numbers is detected, E-node sends a NACK with the sequence numbers of the lost queries back to the sink. The sink retransmits the lost queries after receiving a NACK. When an E-node successfully receives all queries, it checks the P/F bit in received queries. If the P/F bit is set, the E-node sends a ACK to sink. The sink periodically sends message with P/F bit set until the ACK is received.

For upstream reliability, ART proposes a low overhead ACK mechanism. When a new sensing value is obtained, an E-node decides whether the value is an event-alarm. If it is an event-alarm, the E-node sets the \emph{Event Notification} (EN) bit in a new message and saves it into the buffer until the message gets confirmed. Otherwise, the E-node sends a new message to the sink and removes it from the buffer. The sink only sends ACKs to messages with the EN bit set. If an E-node does not receive an ACK for an event-alarm after a timeout, it retransmits the event-alarm message.
 
ART provides congestion control only for upstream flows. The congestion control is handled by E-nodes in a distributed manner. When an event-alarm is sent, the E-node triggers a \emph{congestion timeout} (CTO). CTO is dynamically determined through the round trip time (RTT). If an ACK is not received after the CTO, the E-node broadcasts a \emph{congestion alarm} (CA) message to its neighboring N-nodes. After receiving a CA message, N-nodes will temporarily stop sending their sensing messages. If the E-node still does not receive the ACK after another CTO period, it will increase the broadcast hop-count. If the ACK is received, a congestion-safe message is announced by the E-nodes to resume normal operation on the network.

Table \ref{table:transport_protocols} gives the comparison of the previously discussed transport protocols (Retiring Replicants~\cite{twenty-four} and Burst Forwarding~\cite{twenty-five} are not included in the table, since Retiring Replicants does not provide reliability and Burst Forwarding needs to work with TCP to provide end-to-end reliability and congestion control). Except Hop, all the other transport protocols rely on end-to-end mechanisms to provide end-to-end reliability and congestion control. Hence they are less suitable to ICDT-WSNs. In addition, energy efficiency and buffer management are seldom taken into consideration. Without effective energy efficiency methods and buffer management, the limitation on the sensor node energy and storage can degrade the performance of these protocols in ICDT-WSNs. Therefore, transport protocols for ICDT-WSNs have improvement space in disconnection-oriented design, energy efficiency and buffer management in the future.    

\subsection{Network Layer}
\label{Network Layer}

Routing is a challenging topic in ICDT-WSNs, since traditional routing protocols may not be applicable due to the large requirement of memory for complete link-state information. Existing network protocols can be divided into three categories: single-copy~\cite{twenty-eight} (or forwarding), multiple-copy~\cite{sixteen} (or flooding) and hybrid. \emph{Single-copy protocols} transmit only one copy of a message along a carefully selected path to the destination. \emph{Multiple-copy protocols} transmit several copies of a message to sensor nodes within a network, and expect that at least one copy will reach the destination. \cite{twenty-nine, thirty} point out that multiple-copy protocols, such as epidemic based protocols~\cite{thirty-one}, can improve the delivery capacity. However, Xu and Wang~\cite{thirty-two, thirty-three} have proved that in scenarios where the network connectivity is highly unstable, the buffer occupancy increases as the network size grows. The trade-offs between single-copy and multiple-copy are: 1) multiple-copy protocols can operate with minimal network information, while single-copy protocols need to know more network information to calculate a good path; 2) multiple-copy protocols may cause unnecessary redundancy, but single-copy protocols have to employ expensive packet recovery due to packet loss. 

Network protocols can be grouped into proactive or reactive protocols, depending on when they calculate the routing path. In \emph{proactive network protocols}, all routes are computed and maintained before they are needed. \emph{Reactive network protocols} calculate routes only when needed. To find a good path from a source to a destination is comparatively easier for proactive network protocols, because sensor nodes have a complete view of the network connectivities. But proactive network protocols consume more sensor resources to compute and maintain the routing tables, especially when network topologies change frequently. In reactive protocols, sensor nodes can easily compute and maintain routing tables, since the sizes of routing tables are smaller. However extra delay will be introduced due to the path computation before sending a message. Hybrid network protocols combine both approaches.

In ICDT-WSNs, there are times when a complete path from a source to a destination does not exist. As a result, the network protocols that assume the existence of a complete path from a source to a destination can perform poorly. Because of the limitations on energy and storage of sensor nodes, the network protocols that are not energy efficient and storage friendly are less appropriate for ICDT-WSNs. Zhang gave a comprehensive survey for routing in intermittently connected networks in~\cite{thirty-four} from the perspective of mobile ad hoc networks (MANET) and DTNs. In the rest of this subsection, we outline several recently proposed network protocols for intermittently connected scenarios, and evaluate their network characteristics, node attributes and protocol properties from the perspective of ICDT-WSNs.  

\emph{H + 1 hop}: H + 1 hop~\cite{thirty-one} is a multiple-copy epidemic routing protocol for DTNs. It proposes an approach to reduce the network overhead without sacrificing the delivery speed and delivery rate. H + 1 hop  introduces a threshold called \emph{equilibrium point}, and divides the forwarding process into two sub-processes according to this threshold.  The equilibrium point is the number of duplicate copies of a packet in a network. In epidemic routing it is critical that the amount of infective nodes grows quickly when the number of copies of a packet is low, conversely, the number of infective nodes should decrease as the number of copies of a packet grows. In the first sub-process, H + 1 hop lets nodes distribute the packet to non-infective nodes in the transmission range. After the number of duplicate copies of the packet reaches to the equilibrium point, H + 1 hop enters into the second sub-process, in which nodes that carry a copy of the packet do not distribute to nodes other than the destination.      

\emph{Convergent Hybrid-replication Approach to Routing in Opportunistic Networks (CHARON)}: CHARON~\cite{thirty-five, thirty-six} aims to provide a simple but efficient solution to address the routing problem in highly mobile, sparse sensor networks where the future topologies of networks are unpredictable. It minimizes the number of messages exchanged and provides a way for urgent messages to be delivered quickly.  

CHARON uses \emph{estimated delivery delay (EDD)} as the main routing metric. It forwards messages from nodes with higher EDD to nodes with lower EDD. The EDD of a node $n_j$ is the smallest sum of the $EDD_{n_i}$ and $ICT_{n_i}$ among its neighbors (Equation~\ref{eq: edd}), where $EDD_{n_i}$ is the EDD of its neighbor $n_i$, $ICT_{n_i}$ is the \emph{inter-contact time} between $n_j$ and $n_i$ and $K$ is the set of neighbors of $n_j$.  

\begin{equation} 
EDD_{n_j} = min\{EDD_{n_i} + ICT_{n_i}\}, \forall n_i \in K
\label{eq: edd}
\end{equation} 

CHARON is an energy and buffer space aware protocol. Other than EDD, CHARON employs a customizable multi-variable utility function to calculate routing scores, which includes battery level, free buffer space and other application-specific ones combined in any number of ways.  In order to save energy and reduce the opportunity of a routing loop, messages are forwarded to nodes with lower EDDs and higher utility scores than the one holding the message.

CHARON is a hybrid network protocol. Nodes forward non-critical messages while keeping a local copy of the messages. The local copies are called \emph{Zombie}s, which cost no extra energy, but do utilize node memory, and can only be directly delivered to the sink. Whenever a regular message is received by the sink, its corresponding zombie is removed. With urgent messages, a flooding mode similar to PROPHET~\cite{thirty-seven} is used to make messages arrive at the sink quickly. With high-priority messages, EDD is the only metric used in path calculation in order to minimize the latency. With low-priority messages, both EDD and the utility score are used in order to extend the network lifetime.

\emph{Replication-Based Efficient Data Delivery Scheme (RED) and Message Fault Tolerance-Based Adaptive Data Delivery Scheme (FAD)}:~\cite{thirty-eight} provides two data delivery schemes for delay/fault-tolerant mobile sensor networks (DFT-MSN). Both schemes provide data transmission and buffer management mechanisms. 

Data transmission in RED is based on \emph{nodal delivery probability}, which indicates the likelihood that a node can deliver a data message to the sink. Let $\xi_i$ denote the delivery probability of node $i$, which is initialized to 0. Each sensor node maintains a timer $\Delta$ to calculate its $\xi$. If node $i$ could not transmit any data message during $\Delta$, $\xi_i$ is reduced. If node $i$ transmits a data message to another node $k$, $\xi_i$ is updated according to $\xi_k$. Data messages are stored in a FIFO queue at each node. When node $i$ has a message to send and moves into the communication range of a set of nodes, it earns the delivery probability and available buffer space of each node via simple handshake messages. Then node $i$ transmits the message to neighbor $j$ with available buffer space, the highest delivery probability in the set of neighbor which is also higher than the node's own delivery probability, $\xi_j > \xi_i$. 

Message management in RED uses an erasure-coding approach to address the trade-off between delivery ratio/delay and overhead. The table of delivery probability $p$ and corresponding optimal number of blocks $b$ and minimum replication overhead $S$ is stored in each source node. Whenever a source node generates a message, it checks the table to find the optimal $b$ for its current delivery probability $\xi$ ($\xi = p$), and encodes the data messages into $S \times b$ data blocks, which are then put into the queue for transmission.

FAD is an advanced version of RED that avoids the inaccuracy of the erasure-coding approach used in RED. It takes advantage of both the nodal delivery probability introduced in RED and the message fault tolerance that indicates the redundancy of a message in the network as well as message priority. Unlike RED, nodes that employ FAD keep a copy of the message after transmitting it to another node, which leads to multiple copies of the message and redundancy in the network.  FAD assumes that each message carries a field to keep its fault tolerant value, which is defined as the probability that at least one copy of the message is delivered to the sink by other nodes in the network. Let $F^j_i$ denote the fault tolerant value of message $j$ in the queue of node $i$. $F^j_i$ is initialized to zero when a message is generated. After a node $i$ broadcasts a message $j$ to its neighbor nodes,  there are $Z + 1$ copies of message $j$ in the network, where $Z$ is the set of neighbor nodes of node $i$. The fault tolerance value of the message transmitted to neighbor node $\psi_z$ is updated according to Equation~\ref{eq: fault_tolerance_j}.

\begin{equation} 
F^j_{\psi_z} = 1 - (1 - [F^j_i])(1 - \xi_i) \prod_{m=1, m \neq z}^{Z}(1 - \xi_{\psi_m}).
\label{eq: fault_tolerance_j}
\end{equation} 

\noindent The fault tolerance value of the message at node $i$ is updated according to Equation~\ref{eq: fault_tolerance_i}.

\begin{equation} 
F^j_i = 1 - (1 - [F^j_i]) \prod^{Z}_{m=1} (1 - \xi_{\psi_m}).
\label{eq: fault_tolerance_i}
\end{equation} 

\noindent In Equation~\ref{eq: fault_tolerance_j} and Equation~\ref{eq: fault_tolerance_i}, $[F^j_i]$ is the fault tolerance of the message $j$ in the queue of node $i$ before being transmitted.

Fault tolerance value is used to manage storage. The higher the fault tolerance value is, the less important the message is, since more copies of the message are in the network. To decide whether to store an arriving message $M$ at a node follows two rules: 1) if the queue of a node is full and the fault tolerance of the message at the end of the queue is larger than the fault tolerance of $M$, $M$ is dropped. Otherwise, the message at the end of the queue is dropped, and $M$ is inserted into an appropriate position in the queue according to its fault tolerance; 2) If the fault tolerance of $M$ exceeds a threshold, $M$ is dropped even if the queue is not full.

Data transmission in FAD is based on the delivery probability introduced in RED. When node $i$ has a message $j$ at the front of its queue ready to transmit, and is within the communication range of a set of nodes $Z$, node $i$ multicasts message $j$ to all nodes in $Z$ that have higher delivery probabilities than node $i$ and available buffer space. Meanwhile, node $i$ sets the fault tolerance of message $j$ to just below the threshold in order to reduce unnecessary transmission overhead.  

\emph{Shortest path routing protocol}:~\cite{thirty-nine} proposes a shortest path network protocol for DTNs, which is basically a link-state protocol in which each node has a complete view of the network topology. It assumes that the packet loss is mainly due to buffer overflow, and the waiting time for connections is the main factor of end-to-end delay. Shortest path routing protocol is based on traditional network routing, such as OSPF,  but modifies the decision making strategy for a delay-tolerant environment. 

The shortest path routing protocol uses per-contact routing, which re-computes routing tables when a contact arrives. In order to improve performance, it reduces the time for waiting for a better connection by using \emph{short circuiting}. Short circuiting temporarily assigns an available contact a cost of zero in nodes' local routing tables whenever a contact becomes available. This temporary value is only used to compute the shortest route, but not propagated to other nodes. The combination of per-contact and short circuiting guarantees routing decisions are made with the most recent information and employs serendipitous contact. When making a new connection, nodes exchange summary vectors that list link-state tables received by nodes. Each link-state table has a sequence number, so that nodes can identify the most recent table. After exchanging missing updates, nodes re-compute routing tables and forward messages to other nodes. The \emph{Minimum Estimated Expected Delay (MEED)}, as shown in Equation~\ref{eq: meed}, is the metric used in this protocol to assign a cost to each contact. 

\begin{equation} 
\emph{MEED} = \frac{\displaystyle\sum\limits_{i=1}^{n}d^2_i}{2t},
\label{eq: meed}
\end{equation}   

\noindent where $n$ is the total number of disconnected periods, $d_i$ is the duration of a given disconnected period, and $t$ is the total time interval to observe these disconnections. 

\emph{Spray and Focus}: Spray and Focus~\cite{forty} is a hybrid network protocol for intermittent connected networks. The protocol follows the basic idea of \emph{Spray and Wait}~\cite{forty-one} but improves the \emph{Wait phase} by letting each relay forward its copy further. It assumes that nodes move slowly and periodically transmit beacons to recognize each other's presence. Each node maintains a summary vector, a set of timers and a set of utility values. A summary vector lists message IDs a node has stored and relayed. A timer records the time since two nodes last saw each other. A utility value indicates the possibility a node can deliver a message to another node.

In \emph{Spray phase}, the source node creates $L$ \emph{forwarding tokens} for a message it generates. When two nodes meet, they exchange their summary vectors and check for common messages. If a node $A$ carrying a message copy with forwarding tokens $n > 1$ encounters a node $B$ with no copy of this message, $A$ spawns and forwards a copy of the message with $\lfloor n/2 \rfloor$ tokens to $B$, and reduces the tokens of its own copy to $\lceil n/2 \rceil$. Initially, $n = L$. But if a node has a message copy with only one token, it switches to Focus phase.

In \emph{Focus phase}, the forwarding decision is according to the node's utility. A node $A$ forwards to node $B$ a message destined for node $D$, if and only if Equation~\ref{eq: focuse_phase} is true.

\begin{equation} 
U_B(D) > U_A(D) + U_{th},
\label{eq: focuse_phase}
\end{equation} 

\noindent where $U_B(D)$ is the utility value of B for D, $U_A(D)$ is the utility value of A for D, and $U_{th}$ is a utility threshold. The timer is one possible utility metric. Let $\tau_i(j)$ denote the timer of node $i$ for node $j$ in the network. Initially, set the timer according to Equation~\ref{eq: initially}.

\begin{equation} 
\forall i, j: \tau_i(i) = 0, \tau_i(j) = \infty
\label{eq: initially}
\end{equation} 
Whenever $i$ encounters $j$, update the timer according to Equation~\ref{eq: encounter}. 
\begin{equation} 
\tau_i(j) = \tau_j(i) = 0
\label{eq: encounter}
\end{equation}
At every clock tick, the timer is increased by one (Equation~\ref{eq: every_tick}).
\begin{equation} 
\tau_i(j) = \tau_i(j) + 1,\tau_j(i) = \tau_j(i) + 1.
\label{eq: every_tick}
\end{equation}

According to the forwarding rule in the focus phase, when node $A$ encounters node $B$ at distance $d_{AB}$, it can decide whether $B$ is the next node to forward a message copy according to Equation~\ref{eq: forward},
\begin{equation} 
\forall j \neq B: \tau_B(j) < \tau_A(j) - t_m(d_{AB}),
\label{eq: forward}
\end{equation}
where $t_m(d_{AB})$ denotes the expected time node $A$ takes to move to node $B$ under a given mobility model $m$. 

Each message carries a time-to-live value. If the value expires, the message is removed from the node buffer and its record is also deleted from the node's summary vector.

\emph{TTL-based routing (TBR)}: TBR~\cite{forty-two} aims at maximizing the delivery ratio with minimum network overhead in mobile opportunistic networks through a hybrid scheme. The \emph{time-to-live} (TTL) of a message is used to manage a node buffer and  indicates the delivery priority of messages. 

In TBR, each message has a replication count field $(L_k)$, a TTL field and a list of \emph{previously visited nodes} (LVN). The replication count field indicates how many copies of this message a node can spray. The $L_k$ is set to a user-defined algorithm parameter $L$ initially, and updated after each successful message transmission. The TTL field specifies the valid time (in minutes) of the message in the network. The LVN is used to avoid routing loop. A message is only forwarded to neighbors that are not in the message's LVN.

\begin{landscape}
\begin{table*}[htbp]
\centering
\caption{Comparison of disconnection-oriented routing protocols}
\begin{tabular}{|p{2.2cm}|p{1.8 cm}|p{2.0cm}|p{2.3cm}|p{1.5cm}|p{2.3cm}|p{2.3cm}|p{2.0cm}|}
\hline
  & \textbf{H + 1 hop} & \textbf{CHARON} & \textbf{RED} & \textbf{FAD} & \textbf{Shortest Path Routing Protocol} & \textbf{Spray and Focus} & \textbf{TBR}\\ [0.5ex]
\hline
\hline
\textbf{Network Model} & DTN & Low density opportunistic WSNs &  DFT-MSN & DFT-MSN & DTN & Intermittently connected mobile networks & Mobile opportunistic networks\\ [0.5ex]
\hline
\textbf{Node Mobility} & Normal & High mobility & Normal & Normal & - & Slow mobility & Normal \\[0.5ex] 
\hline
\textbf{Energy Aware} & $\times$ & $\checkmark$ & $\times$ & $\times$ & $\times $ & $\times$ & $\times$ \\[0.5ex]
\hline
\textbf{Buffer Aware} & $\times$ & $\checkmark$ & $\checkmark$ & $\checkmark$ & $\times$ & $\times$ & $\checkmark$ \\[0.5ex]
\hline
\textbf{Buffer Management} & $\times$ & $\times$ & $\checkmark$ & $\checkmark$ & $\times$ & $\times$ & $\checkmark$ \\[0.5ex]
\hline
\textbf{Buffer Size} & Sufficient & Limited & Limited & Limited & Sufficient & Sufficient & Limited \\[0.5ex]
\hline
\textbf{Computation Complexity} & Simple computation & Simple computation & Requires extra computation for decoding & Simple computation & Computation and exchange & Simple computation & Simple computation \\[0.5ex]
\hline
\textbf{Single-copy case/ Multiple-copy case/ Hybrid} & Multiple-copy &Multiple-copy (urgent messages) + Hybrid (normal messages) & Single-copy & Multiple-copy & Single-copy & Hybrid & Hybrid\\[0.5ex]
\hline
\end{tabular}
\label{table:network_protocols}
\end{table*}
\end{landscape}

Each node maintains a forward list, a delete list and an acknowledged message list. The \emph{forward list} stores the forwarding priorities of messages. When a contact becomes available, the message on the top of the list will be delivered first. Let $P_{k_f}$ denotes the forwarding priority of the message $m_k$ (Equation~\ref{eq: forward_prob}), 
\begin{equation} 
P_{k_f} = \frac{1}{H_k \times TTL_k \times s_k},
\label{eq: forward_prob}
\end{equation}
where $TTL_k$ is the TTL of $m_k$, $H_k$ is the hop count of $m_k$ and $s_k$ is the size of $m_k$. The \emph{delete list} stores the a prioritized list of messages to be deleted. If a node buffer is full when it receives another message with higher priority, the message at the end of the list will be deleted from the buffer. Let $P_{k_d}$ denote the deleting priority of the message $m_k$ (Equation~\ref{eq: delete_prio}), 
\begin{equation} 
P_{k_d} = \frac{L_k}{s_k},
\label{eq: delete_prio}
\end{equation}

\noindent where $L_k$ is the replication count field value of $m_k$, and $s_k$ is the size of $m_k$.
The \emph{acknowledged message list} records the messages that have been delivered successfully.

When two nodes encounter each other, they first exchange their acknowledged message lists. If nodes have messages in their buffer that are in the acknowledged message lists, the messages are removed. If a node has some messages destined to the other node, it transmits those messages. If a node has forwarding messages, it picks the messages from the top of the forward list as long as the messages meet the conditions $L_k > 1$, $s_k < S_{max}$ and the other node is not in the messages' LVNs. $S_{max}$ is the maximum transmittable message size of a contact. TBR calculates it every time when a new contact is available. The $L_k$ of each message is divided by 2 before forwarding a copy to the other node. When $L_k = 1$, a message can only be delivered to the destination directly or dropped from the buffer when its TTL expires. 

Considerations on energy efficiency, buffer management and computation complexity are necessary and important to network protocols in ICDT-WSNs. Table \ref{table:network_protocols} gives a comparison of the protocols mentioned in this subsection according to these considerations. As can be seen, none of these network protocols meets all requirements of ICDT-WSNs. The reason is that these network protocols are designed for DTNs, opportunistic networks, MANETs and DDTMs~\cite{forty-three}, but not for ICDT-WSNs. Hence, energy efficient network protocols for ICDT-WSNs with buffer management and low computation complexity are a future research direction.

\subsection{Link Layer}
\label{Link Layer}

The communication media of ICDT-WSNs is a wireless channel whose nature is broadcast. This feature makes the medium access and data transmission over the common medium complicated. Moreover, restrictions on nodes' energy, storage and computation capability, asymmetric links, dynamic topology, and the large number of nodes of a network make media access control (MAC) protocols for WSNs distinct from the protocols for traditional networks. 

Due to the power limitation of sensor nodes, energy efficiency is a primary goal in MAC protocol design. The scalability of the network size, topology and node density is a further goal. Because all nodes in a WSN cooperate together to accomplish a mission, the requirements on fairness, latency, throughput and bandwidth utilization are less strict than in traditional networks. 


There are four factors that can drain energy quickly without effective management: collisions, overhearing, overhead and idle listening. Much work on addressing these four issues has been done  over the past decades, which can be divided into four groups: scheduled protocols,  protocols with common active periods, preamble sampling protocols and hybrid protocols that combine the techniques used in the former three groups~\cite{forty-four}.  

In \emph{scheduled protocols}, all nodes in the network follow a well-designed schedule to communicate. Time division multiple access (TDMA) is a canonical module. A well-designed schedule can avoid collisions and overhearing, and can minimize idle listening. However, to design such a schedule for a large WSN is overly complex. Additionally, the scalability and flexibility of scheduled protocols are limited in scheduled protocols.

In \emph{common active period protocols}, sensor nodes have common active periods to communicate. So synchronization is required in this category of protocols. These protocols reduce the energy spent in idle listening, but bring up the problem on the suitable length of active slots and the problem of sleep delay. Even though sleep delay is not a significant problem in ICDT-WSNs, the length of the active slot is tightly relevant to energy efficiency.

\emph{Preamble sampling protocols} reduce the energy spent on synchronization by using long preambles, but the energy consumption is transferred from receivers to transmitters. Receivers only need to wake up for a short period of time to sense the channel, but transmitters need to spend more energy transmitting long preambles. This is acceptable when the traffic is light; otherwise, long preamble transmitting will consume too much energy. Moreover, if collisions happen, the energy spent on long preambles becomes increasingly problematic. The duty cycle is also limited in this type of MAC protocols, since the length of preamble is relevant to the channel check interval of nodes.

Most of the MAC protocols for WSNs are suitable for ICDT-WSNs when the propagation delay in networks is normal (e.g., RF communication in terrestrial environment). However in the scenario where the propagation delay is very long (e.g., acoustic communication in underwater environment), propagation-delay tolerant MAC protocols are required~\cite{forty-five, forty-six}. In this subsection, we focus our study on MAC protocols designed for the scenarios where propagation delay is not a significant problem. We outline several protocols that are well known or recently designed in each category, and study their advantages and disadvantages. 

\emph{Traffic-adaptive medium access protocol (TRAMA)}: TRAMA~\cite{forty-seven} assumes there is a single, time-slotted channel for both signaling and data transmission, and adequate synchronization is attained. Time is organized as signaling slots and transmission slots. Signaling slots are random-access, and transmission slots are scheduled-access. TRAMA consists of three components: the Neighbor Protocol (NP), the Schedule Exchange Protocol (SEP) and the Adaptive Election Algorithm (AEA). TRAMA starts in signaling slots in which nodes use NP to obtain two-hop neighborhood information. Then during transmission slots, nodes use SEP to build and maintain traffic-based schedules for the one-hop neighbors before deciding their state (transmit, receive or sleep) by AEA. AEA selects collision-free transmitters and receivers based on the information from NP and SEP. In this way, nodes sleep until they need to transmit or receive in the transmission slots.  

\emph{Multi-Channel Lightweight Medium Access Control (MC-LMAC)}: MC-LMAC \cite{forty-eight} is a multi-channel access scheduled MAC protocol, which aims at improving the achievable throughput in WSNs rather than energy efficiency. In MC-LMAC, each node chooses one time slot from a channel in a distributed way, which guarantees that the same slot/channel pair is not used by more than one node which avoids conflicting transmissions.   

Nodes in MC-LMAC transit between five states: initialization, synchronization, discovery, time-slotted channel selection, and medium access. When nodes join the network, they begin in the initialization state in which nodes sample the medium for incoming packets to synchronize with the network. If such packets are received, nodes move into the synchronization state and synchronize with the network by the current time slot information and frame number carried in the packets. Then, nodes follow the schedule to receive packets in the upcoming slots. If nodes have data to send, they pick up a random wake-up frame in channels. Before the wake-up frames, nodes enter the discovery state. In this state, each node gets a list of free time-slots and channels in neighborhood. In the list, each entry represents one channel, which is a string of 1s or 0s that indicates whether the slot is used or free with the length of the string equaling the number of time-slots in that channel. After the discovery state, nodes come into the time-slotted channel selection state. In this state, every node executes bitwise OR operation against the list to pick a free time-slot channel pair that is not used by the node's one-hop neighbors. If a node successfully chooses an empty time-slot, the node proceeds into the media access state, otherwise, it goes back to the synchronization state. In the media access state, nodes can transmit data in their selected time slots of the selected channels. During other slots, nodes detect potential incoming packets. If collisions are detected, the nodes go back to the synchronization state. If a synchronization error happens, the nodes go back to the initialization state.     

\emph{S-MAC}: S-MAC~\cite{forty-nine} assumes that applications have long idle periods, and can tolerate some latency. S-MAC reduces idle listening by letting nodes sleep for a fixed period and then wake up for the same duration to listen for any node that wants to talk to it. In order to reduce control overhead, neighboring nodes need to synchronize to have the same listen/sleep schedule. In this way, each node maintains a schedule table, which stores the schedules of all its neighbors. In the beginning, nodes need to build the schedule tables. First, every node listens for a period of time. If no schedule is received during this time, it randomly chooses a time to go to sleep and broadcasts a SYNC message with its schedule to inform that it will go to sleep after $t$ seconds. If a schedule is detected before the node chooses its own schedule, the node sets its schedule to the received schedule, and rebroadcasts the schedule after a random delay $t_d$ to inform that it will go to sleep after $t - t_d$ seconds. If a node receives a different schedule after it has set and broadcasted its own schedule, it adopts both schedules. The listening duration is divided into two parts. The first part is for nodes to synchronize to have the same listen/sleep schedule through SYNC packets. Nodes that receive SYNC packets adjust their timers right away. The second part is for receiving RTS (request to send) packets. Both parts are divided into several time slots for senders to detect collisions (carrier sense).

In order to avoid collisions, S-MAC sets a duration field in each transmitted packet to indicate how long the remaining transmission will be. If a node receives a packet that is not destined for it, the node records the duration field value called \emph{network allocation vector} (NAV) and sets a timer. NAV indicates how long the receiving node should keep silent. Once the timer starts, the NAV value is decreased until it reaches to zero. When a node has data to send, it first checks the current NAV value. If the NAV value is zero, the node performs carrier sense. Broadcast packets are sent without using RTS/CTS (clear to send) packets. Unicast packets are sent after exchanging RTS/CTS. After a receiver receives a unicast packet, it sends ACK back to the sender. If a node fails to acquire access to the medium, it goes back to sleep until the receiver is free and listens again.

To avoid overhearing, S-MAC let nodes that hear the RTS or CTS packets sleep until the current transmission is over. When a node receives a packet destined for other nodes, it updates its NAV by the duration field in the packet.

S-MAC reduces the resources spent on retransmitting a long data packet by divided the packet into many small fragments and transmits them in a burst. A sender waits for ACK from its receivers for each fragment. If an ACK is not received, the sender extends the reserved transmission time for one more fragment, and retransmits the packet.

\emph{Timeout-MAC (T-MAC)}: T-MAC~\cite{fifty} follows the basic idea of S-MAC with the active/sleep schedule and the way nodes decide their schedules. However, instead of using the fixed duty cycle in S-MAC, T-MAC introduces an adaptive duty cycle to improve energy efficiency. In T-MAC, nodes periodically wake up to communicate with their neighbors, and then go back to sleep until the next frame. During the sleep time, new generated messages are queued. The communication between nodes follows the sequence of RTS/CTS/DATA/ACK similarly to S-MAC. In the active period, nodes keep listening and potentially transmit. When the active period  ends, nodes go back to sleep. If no active event occurs for a time $TA$, the active period ends, which is called the adaptive duty cycle. In order to optimize the sleep period, T-MAC moves all communication to the beginning of the active time in a burst. The minimum length of $TA$ should be larger than the maximum contention duration and the time to exchange RTS and CTS.  The length of $TA$ is a key factor in energy consumption. The longer $TA$ is, the more energy is consumed. By reducing the active duration, T-MAC can achieve better energy efficiency. 

The adaptive duty cycle can cause an early sleep problem in which a node goes to sleep when its neighbor still has data to send to it. To solve this problem, T-MAC uses the \emph{future-request-to-send} (FRTS) packet, which indicates how long a node needs to be awake. A FRTS packet is sent right after a node overhears a CTS message destined for another node. When a node receives a FRTS, it will stay active for the time indicated by the FRTS rather than going back to sleep. Then its neighbor can send data to it after the channel is clear. This feature increases the throughput of T-MAC by introducing additional overhead.  

\emph{B-MAC}: B-MAC~\cite{fifty-one} is a preamble sampling protocol that searches for outliers during channel arbitration. If there exists an outlier from the received signals during the sampling period, B-MAC declares the channel is clear. If five samples are taken and no outlier is detected, the channel is busy. So a good estimation of noise floor is important in this procedure. In B-MAC, each node takes signal strength samples at a time when the channel is assumed to be clear, such as immediately after transmitting a packet. The average value of the samples is used as a simple low pass filter for the noise floor estimate.

During the \emph{low power listening} (LPL) state, a node wakes up and turns on its radio to check activity. If no activity is detected, the node is forced back to sleep after a timeout. If activity is detected, the node turns power on and stays awake for the time required to finish receiving the incoming packet. After reception, the node goes back to sleep. The length of preamble should be at least the interval that the channel is in LPL state.  

\emph{X-MAC}: X-MAC~\cite{fifty-two} ameliorates the overhearing problem by dividing one long preamble into a series of short preamble packets with a small gap between every two packets, during which the transmitter pauses transmitting. Each preamble packet contains the ID of the target node. When a node wakes up and receives a preamble packet, it looks up the target ID in the packet. If the node is not the target, it goes back to sleep immediately. If it is the target, it sends an acknowledgement to the transmitter during the gap between the preamble packets. After the transmitter receives an acknowledgement, it stops transmitting preamble and starts sending the data packet. Similar to B-MAC, the length of the preamble sequence must be greater than the maximum sleep period of receivers.

When multiple transmitters want to send packets to the same node, there is no need to send preambles to the node that is already awake. X-MAC addresses this problem by letting the receiver keep awake for a short period of time after receiving a data packet. During this period, if a transmitter hears the acknowledgement from its target node while waiting for a clear channel, it waits for a random back-off time and then transmits without sending any preamble. The back-off time should be long enough to allow the on-going transmission to finish, and the active time of the receiver after receiving a data packet should be equal to the longest back-off time.

\begin{landscape}
\begin{table*}[htbp]
\centering
\small
\caption{Comparison of MAC protocols for WSNs}
\begin{tabular}{|p{2cm}|p{2.5cm}|p{6.5cm}|p{6.5cm}|}
\hline
\textbf{Protocol} & \textbf{Mode} & \textbf{Advantages} & \textbf{Disadvantages} \\ [0.5ex]
\hline
\hline
\multirow{2}{*}{\textbf{TRAMA}} & Schedule based & $\bullet$ Automatically adapt scheduling to traffic load.  & $\bullet$ Computation complexity is high. \\ & & $\bullet$ Nodes wake up only when they have data to send or receive & $\bullet$ Assume adequate synchronization among nodes. \\ [0.5ex]
\hline
\multirow{2}{*}{\textbf{MC-LMAC}} & Schedule based & $\bullet$ Provide high throughput  & $\bullet$ Low energy efficiency \\ && $\bullet$ Low computation complexity. & $\bullet$ Multiple channels required. \\[0.5ex] 
\hline
\multirow{4}{*}{\textbf{S-MAC}} & Common active & $\bullet$ Set up a common active period to reduce overhead. & $\bullet$ Applications must tolerate some latency. \\ &period& $\bullet$ Reduce overhearing in unicast. & $\bullet$ Fixed duty cycle. \\ && $\bullet$ Reduce energy spent in retransmission by dividing a long data packet into small fragments. &\\ && $\bullet$ Comparatively less computation. & \\[0.5ex]
\hline
\textbf{T-MAC} & Common active period & $\bullet$ Adaptive duty cycle to reduce energy consuming. & $\bullet$ Additional overhead needed to support adaptive duty cycle. \\[0.5ex]
\hline
\textbf{B-MAC} & Preamble based & $\bullet$ Low computation complexity. & $\bullet$ Long preamble causes unnecessary overhearing. \\&&$\bullet$ No synchronization needed. & \\[0.5ex]
\hline
\textbf{X-MAC} & Preamble based & $\bullet$ Reduce unnecessary overhearing by dividing long preamble into a series of small preamble packets. & $\bullet$ Introduce new overheads. \\&& $\bullet$ Keep nodes awake after finishing receiving data to avoid sending preambles to awake nodes. & \\[0.5ex]
\hline
\textbf{Z-MAC} & Hybrid (CSMA/TDMA) & $\bullet$ Increase channel utilization. & $\bullet$ Not easy to introduce new nodes, since time slots are assigned to nodes when the network is developed. \\  &&$\bullet$ No extra overhead for collision avoidance after initial set up phase.& $\bullet$ Comparatively high computation complexity during the initial set up phase. \\[0.5ex]
\hline
\end{tabular}
\label{table:mac_protocols}
\end{table*}
\end{landscape}

\emph{Zebra MAC (Z-MAC)}: In Z-MAC~\cite{fifty-three}, carrier sense multiple access (CSMA) is used inside a large time division multiple access (TDMA) slot, such that Z-MAC can utilize CSMA when the traffic is light and switch to TDMA if the traffic becomes heavy. Initially, each node runs the DRAND~\cite{fifty-four} algorithm to be assigned a time slot that is unique within the node's two-hop neighborhood. In \emph{low contention level} (LCL), a node can compete for any time slot. However, in \emph{high contention level} (HCL), a node can compete for a slot only when it is the owner of the slot or a one-hop neighbor of the owner of a slot. When a node has data to transmit, it always performs carrier sense first and transmits after the channel is clear. If the node is the owner of a slot, it has the highest priority to send. Otherwise, it needs to wait a random back-off time, after which the node can transmit in that slot if the slot is still unused. The back-off time needs to be long enough in order to allow the owner access to the slot. 

A node is in the HCL when it receives an \emph{explicit contention notification} (ECN) message from a two-hop neighbor within the last $t_{ECN}$ period. Otherwise, it is in LCL. ECN messages are sent by a transmitter when it detects the channel is in high contention by measure the noise level of the channel. When a node receives an ECN message, it sets a local HCL flag. The HCL flag is automatically reset unless the node receives another ECN message within $t_{ECN}$. The system sets the refresh time $t_{ECN}$.
 
As we can see from table \ref{table:mac_protocols}, each link-layer protocol has its own benefits and drawbacks. Most of the protocols are either suitable for light contention scenarios or heavy contention scenarios. While those protocols suitable for both scenarios have high computation complexity. For applications whose traffic is predictable, most existing protocols can be applied. However, for those unpredictable situations, protocols that are without high computational complexity and can handle different scenarios and  are preferable.

\section{Open Problems and Research Directions}
\label{Open Problems}  

In Table~\ref{table:protocols_summary}, we check the communication protocols mentioned in the previous section against the requirements of ICDT-WSNs. From the table, we can see that seldom protocols meet all requirements,  since the existing communication protocols are not designed for ICDT-WSNs. As a result, there is a lot of space for further research. In this section, we outline some of the open problems in ICDT-WSNs and suggest the research directions in addressing these problems.

\subsection{Open Problems and Research Directions in Transport Protocols}
\label{Open Problems in Transport Protocols}

In Subsection \ref{Transport Layer}, we discuss the existing transport protocols, and compare several protocols designed in recent years from reliability, congestion control, energy efficiency and buffer management in Table \ref{table:transport_protocols}. Due to the intermittent connection of ICDT-WSNs, the hop-by-hop transport protocols that provide end-to-end reliability and congestion control through hop-by-hop manner are more appropriate. However, existing hop-by-hop transport protocols do not take energy efficiency and buffer management into account. As a result, energy efficiency and buffer management become the open problems in transport protocols.

In order to be energy efficient and storage friendly, an effective buffer management mechanism in a transport protocol can save space for more important messages, and can reduce energy consumption in unnecessary retransmissions. Better-designed transport protocols can take better advantage of the information from the control messages to improve reliability, congestion control and buffer management. As a result, the energy consumption can be reduced by reducing the volume of the control message traffic.

Recently proposed hop-by-hop transport protocol~\cite{fifty-five}, \emph{Acksis}, providing both reliability and congestion control, addressed the buffer management problem by taking advantage of the per-block based acknowledgement. Through this mechanism, Acksis can enhance the network delivery rate by employing in-network storage regardless of the traffic situations of ICDT-WSNs.

 \begin{table*}[htbp]
\centering
\small
\caption{Comparison of communication protocols with ICDT-WSN requirements}
\begin{tabular}{|p{3.0cm}|p{2.2cm}|p{1.6cm}|p{1.4cm}|p{2cm}|}
\hline
\textbf{Protocol} & \textbf{Disconnection-oriented} & \textbf{Energy Efficiency} & \textbf{Storage Friendly} & \textbf{Computation Simplicity}\\ [0.5ex]
\hline
\hline
\textbf{Hop} & $\checkmark$ & $\checkmark$ & $\times$ & $\checkmark$ \\[0.5ex]
\hline
\textbf{RCRT} & $\times$ & $\times$ & $\times$ & $\checkmark$ \\[0.5ex]
\hline
\textbf{STCP} & $\times$ & $\times$ & $\checkmark$ & $\checkmark$ \\[0.5ex]
\hline
\textbf{Flush} & $\times$ & $\times$ & $\times$ & $\checkmark$ \\[0.5ex]
\hline
\textbf{ART} & $\times$ & $\checkmark$ & $\times$ & $\times$ \\[0.5ex]
\hline
\hline
\textbf{H + 1 hop} & $\checkmark$ & $\times $ & $\times$ & $\checkmark$ \\[0.5ex]
\hline
\textbf{CHARON} & $\checkmark$ & $\checkmark $ & $\times$ & $\checkmark$ \\[0.5ex]
\hline
\textbf{RED} & $\checkmark$ & $\times$ & $\checkmark$ & $\times$ \\[0.5ex]
\hline
\textbf{FAD} & $\checkmark$ & $\times$ & $\times$ & $\checkmark$ \\[0.5ex]
\hline
\textbf{Shortest Path Routing Protocol} & $\checkmark$ & $\times$ & $\times$ & $\times$ \\[0.5ex]
\hline
\textbf{Spray and Focus} & $\checkmark$ & $\times$ & $\times$ & $\checkmark$ \\[0.5ex]
\hline
\textbf{TBR} & $\checkmark$ & $\times$ & $\checkmark$ & $\checkmark$ \\[0.5ex]
\hline
\hline
\textbf{TRAMA} & N/A & $\checkmark$ & N/A & $\times$\\[0.5ex]
\hline
\textbf{MC-LMAC} & N/A & $\times$ & N/A & $\checkmark$ \\[0.5ex]
\hline
\textbf{S-MAC} & N/A & $\checkmark$ & N/A & $\checkmark$\\[0.5ex]
\hline
\textbf{T-MAC} & N/A & $\checkmark$ & N/A & $\checkmark$ \\[0.5ex]
\hline
\textbf{B-MAC} & N/A & $\checkmark$ & N/A & $\checkmark$\\[0.5ex]
\hline
\textbf{X-MAC} & N/A & $\checkmark$ & N/A & $\checkmark$ \\[0.5ex]
\hline
\textbf{Z-MAC} & N/A & $\checkmark$ & N/A & $\times$ \\[0.5ex]
\hline
\end{tabular}
\label{table:protocols_summary}
\end{table*}

\subsection{Open Problems and Research Directions in Network Protocols}
\label{Open Problems in Network Protocols}

Subsection \ref{Network Layer} introduces the categories of existing network protocols. Table \ref{table:network_protocols} gives the comparison for several network protocols designed in recent years from network characteristics, node attributes and protocol properties. Reactive protocols are more suitable for ICDT-WSNs due to the intermittent connectivity. Reactive protocols for ICDT-WSNs should also consider the energy efficiency, buffer management and computation complexity. There are several reactive protocols with simple computation complexity, however, only a few of them take energy efficiency and buffer management into consideration. Hence, energy efficiency coupled with buffer management is also an open problem in network layer protocols. 

This paper argues that hop-by-hop transport protocols are more appropriate to ICDT-WSNs, hence, the cross-layer protocols that combine reactive network protocols with hop-by-hop transport protocols is an approach for the open problems of transport and network layers. Employing such a cross-layer protocol can introduce less overhead than employing a transport protocol and a network protocol. As a result, this type of cross-layer protocols can be more energy efficient. In addition, the storage management mechanism of this type of cross-layer protocol can address the insufficient storage problem in both transport layer and network layer. 

Recently proposed scheme SMITE~\cite{fifty-six} addressed the storage limitation of nodes by applying \emph{Bloom filter}. Bloom filter can improve the storage capability of nodes with low computation complexity. However, it introduces the complexity to higher layer when providing reliability and integrity. Diff-Max~\cite{fifty-seven}, another recent work, proposed a new framework to separate routing and scheduling in backpressure algorithm for better throughput, easier implementation and flexibility. Meanwhile, Ji at al.~\cite{fifty-eight} have explored the inefficiencies of backpressure. As a result, the backpressure algorithm can lead to energy inefficiency. These recently proposed works address the open problems in network protocols in ICDT-WSNs through either introducing new schemes or optimizing the existing mechanisms to be more energy efficient or storage friendly. However, these approaches do not fully address the open problems and leave space open for future research.

\subsection{Open Problems and Research Directions in MAC Protocols}
\label{Open Problems in MAC Protocols}

The MAC protocols are discussed in Subsection \ref{Link Layer}.  Most of the existing MAC protocols are traffic-dependent efficient. The protocols that are efficient in both heavy and light traffic are with high computation complexity. Accordingly, the protocols with low computation complexity that are traffic-independent efficient is an open problem to further research. 

Preamble based protocols can save energy spent on synchronization when traffic is light, while common active period protocols can save energy spent on preamble sending when traffic is heavy. Both categories have better scalability and less computation complexity than scheduled protocols. Thus, the combination of common active period protocols and preamble based protocols with low computational complexity is a way to handle the scenarios where the traffic is unpredictable. 
 
\section{Conclusion}
\label{Conclusion}

This article provides a survey of communication protocols from the perspective of Intermittently Connected Delay-Tolerant Wireless Sensor Networks (ICDT-WSNs) --- a branch of WSNs with all the limitations from WSNs and the intermittent connectivity of DTNs. This article introduces the key attributes of ICDT-WSNs and points out that the existing communication protocols are unsuitable for such networks. In order to develop the direction of improvement on the communication protocols for ICDT-WSNs, we study several protocols that have been designed in recent years, from transport layer to link layer, from the perspective of ICDT-WSNs. We also list out the open questions in communication protocols for ICDT-WSNs, and provide possible solutions for these problems.


Out of the broad range of transport layer and network layer protocol, hop-by-hop transport protocols and reactive network protocols are appropriate in ICDT-WSNs domain. However, the existing transport and network layer protocols are either designed for WSNs or DTNs, so that the storage and energy limitations of the nodes are not sufficiently taken into account. The existing link layer protocols are suitable to ICDT-WSNs where the traffic load is predictable. But for the scenarios where the traffic load is unpredictable, the existing link layer protocols haven't provided methods with simple computation complexity and good scalability. In addition, the works for the scenarios where the propagation delay is a significant problem are insufficient.
 



\end{document}